\documentclass[fleqn,twoside]{article}
\usepackage{espcrc2}
\usepackage{epsfig}
\usepackage{graphicx}
\newcommand{\beq}{\begin{equation}}
\newcommand{\eeq}{\end{equation}}
\newcommand{\be}{\begin{eqnarray}}
\newcommand{\ee}{\end{eqnarray}}
\title{Neutrino-nucleus cross section in the impulse approximation regime}

\author{Omar Benhar\address[INFN]{INFN, Sezione di Roma \\ Piazzale Aldo Moro, 2.
        I-00185 Roma, Italy}\address[DF]{Dipartimento di Fisica, Universit\`a ``La Sapienza" \\
        Piazzale Aldo Moro, 2.  I-00185 Roma, Italy}, % 
        Nicola Farina\addressmark[DF]
        }

\begin{document}

\begin{abstract}
In the impulse approximation regime the nuclear response to a weakly interacting probe
can be written in terms of the measured nucleon structure fuctions and the target
spectral function, yielding the energy and momentum distribution of the constituent 
nucleons. We discuss a calculation of charged current neutrino-oxygen interactions
in the quasielastic channel, carried out within nuclear many body theory.
The proposed approach, extensively and successfully employed in the
analysys of electron-nucleus scattering data, 
allows for a parameter free prediction of the neutrino-nucleus cross section,
whose quantitative understanding will be
critical to the analysis of the next genaration of high precision neutrino oscillation
experiments.

\vspace{1pc}
\end{abstract}

\maketitle

%%%%%%%%%%%%%%%%%%%%%%%%%%%%%%%%%%%%%%%%%%%%%%%%%%%%%%%%%%%%%%%%%%%%%%%%%%%%%%%%%%%%%%%%%%%%
\section{Introduction}
%%%%%%%%%%%%%%%%%%%%%%%%%%%%%%%%%%%%%%%%%%%%%%%%%%%%%%%%%%%%%%%%%%%%%%%%%%%%%%%%%%%%%%%%%%%%

In experimental searches of oscillations, neutrinos are detected trough their 
interactions with nucleons bound in nuclei (e.g. oxygen). Therefore, the knowledge of the
neutrino-nucleus cross section at a fully quantitative level will be critical to reduce 
the systematic uncertainty of the next generation of high precision measurements. 

Accurate calculations of the weak nuclear response at moderate momentum transfer can be carried
out within nuclear many body theory (NMBT) using nonrelativistic wave functions to
describe the target initial and final states and expanding the current operator in powers
of $(|{\bf q}|/m)$, ${\bf q}$ and $m$ being the momentum transfer and the nucleon 
mass, respectively. Applications of this approach to the analysis of electron scattering 
data has been very successful (for a review see, e.g., ref.\cite{book}).

At the higher values of $|{\bf q}|$ corresponding to energies of the 
beam particles in the few GeV region, the description of the final states in terms 
of nonrelativistic nucleons is no longer
possible, and some simplifying assumptions have to be made in order to take into
account both the presence of relativistic particles
and the occurrence of inelastic processes. 

An approximation scheme widely employed to describe the region of high momentum transfer 
is based on the idea that, as the beam particle probes a region of extension 
$\sim 1/|{\bf q}|$ of the target, at large $|{\bf q}|$ the scattering process involves 
only one nucleon, the remainig A-1 particles
acting as spectators. 

In the simplest implementation of this scheme, usually referred to as
impulse approximation (IA), the hadronic
state produced at the weak interaction vertex is assumed to be totally decoupled from 
the recoiling system, so that the description of its motion reduces to a purely kinematical 
problem. 
On the other hand, the target intial state, as well as the final state of the spectator 
particles, can be safely treated within NMBT using nonrelativistic dynamics.

In Section 2, after giving the expression of the neutrino-nucleus cross section in the 
IA regime, we briefly outline a theoretical approcah, based on NMBT, that allows for both a 
realistic description of the initial state and the inclusion of
corrections arising from final state interactions (FSI). 
The results of calculation for the test case of quasielastic charged current 
interactions with oxygen are presented in Section 3, while Section 4 is devoted 
to a summary and the conclusions.

%%%%%%%%%%%%%%%%%%%%%%%%%%%%%%%%%%%%%%%%%%%%%%%%%%%%%%%%%%%%%%%%%%%%%%%%%%%%%%%%%%%%%%%%%%%%
\section{The neutrino-nucleus cross section in the impulse approximation regime}
%%%%%%%%%%%%%%%%%%%%%%%%%%%%%%%%%%%%%%%%%%%%%%%%%%%%%%%%%%%%%%%%%%%%%%%%%%%%%%%%%%%%%%%%%%%%

The Born approximation cross section of the weak charged current process
\beq
\nu_\ell + A \rightarrow \ell^- + X\ , 
\label{process}
\eeq
where $A$ and $X$ denote the target nucleus and the undetected hadronic final state, respectively, 
can be written in the form
\beq
\frac{d\sigma}{d\Omega_\ell dE_\ell} = \frac{G^2}{32 \pi^2}\ 
\frac{|{\bf k}^\prime|}{|{\bf k}|}\ 
 L_{\mu \nu} W^{\mu \nu}\ .
\label{sigmX:1}
\eeq
In the above equation, $G$ is the Fermi coupling constant, 
$E_\ell$ is the energy of the final state lepton and ${\bf k}$ and 
${\bf k}^\prime$ are the neutrino and charged lepton momentum, respectively. 
The tensor $L_{\mu \nu}$ is fully specified by the kinematical variables of the 
leptons, while $W^{\mu \nu}$ is
defined in terms of the target initial and final states and current 
according to
\be
\nonumber
W^{\mu \nu} & = & \sum_X
\langle A | J^{{\mu^\dagger}} | X \rangle \langle X | J^\nu | A \rangle \\
  &  & \ \ \ \ \ \ \ \ \ \ \ \ \ \ \ \times \delta^{(4)}(P_A+q-P_X) \label{Wmunu:1}\ , 
\label{WmunuA}
\ee
$P_A$, $P_X$ and $q$ being the target initial and final four momenta and the
four momentum transfer, respectively.

The IA scheme outlined in Section 1 is based on the assumptions 
that: i) the target nucleus is seen
by the neutrino as a collection of individual nucleons and ii)
the hadron produced at the weak interaction vertex is unaffected by either 
Pauli blocking or final state interactions (FSI) with the spectator nucleons.

According to this picture, the state $|X \rangle$ appearing in Eq.(\ref{WmunuA}) can 
be written in a simple factorized form and the tensor $W^{\mu \nu}$ of Eq.(\ref{WmunuA}) 
reduces to (see, e.g. ref.\cite{benhar91})
\beq
W^{\mu \nu}_{IA}  = \int d^4 p\ P(p)\
w^{\mu\nu}({\widetilde p},{\widetilde q})\ ,
\label{tensor:IA}
\eeq
where $P(p)$ ($p\equiv(p_0,{\bf p})$) is the nuclear spectral function \cite{benhar89}, 
yielding the probability of removing
a nucleon of momentum ${\bf p}$ from the target leaving the residual system with 
energy $E = m - p_0$.
Eq.(\ref{tensor:IA}) shows that the IA formalism allows one to describe scattering off a
{\it bound} nucleon carrying four momentum $p$ in terms of the tensor $w^{\mu\nu}$, describing 
the weak charged current interactions of a nucleon in free space. 
Binding effects are  
taken care of by replacing the bound nucleon four momentum $p\equiv(p_0,{\bf p})$ with 
${\widetilde p}\equiv(\sqrt{{\bf p}^2+m^2},{\bf p})$ and the 
four momentum tranfer $q\equiv(\nu,{\bf q})$ with ${\widetilde q}= p - {\widetilde p} + q$. 
\cite{benhar91,deforest83,benhar93}.

The IA cross section can be obtained from Eqs.(\ref{sigmX:1}) and (\ref{tensor:IA}) 
using the available parametrizations of the structure functions (both elastic and inelastic)
entering the definition of $w^{\mu\nu}$ and the nuclear spectral functions 
resulting from calculations 
carried out within nuclear many body theory (NMBT) \cite{benhar94}.

NMBT also provides a consistent framework to include corrections to the IA in a systematic
fashion. The effects of FSI, that have long been recognized to be large in both inclusive and 
exclusive electron-nucleus scattering processes (a series of review papers on electron-nucleus
scattering can be found in ref.\cite{book}), can be taken into account within the
approach of ref. \cite{benhar91}, based on the assumptions that i) the 
hadron produced at the weak interaction vertex moves along a straight trajectory 
with constant velocity (eikonal approximation) and ii) the spectator 
nucleons are seen as a collection of fixed scattering centers (frozen spectators approximation). 
The resulting cross section can be written in terms of the IA results according to
\beq
\frac{d\sigma_A}{d\Omega_\ell dE_\ell} = \int dE_\ell^\prime\
f_q(E_\ell - E_\ell^\prime)\
\left(\frac{d\sigma}{d\Omega_\ell dE_\ell^\prime} \right)_{IA}.
\label{folded:sigma}
\eeq
The above equation shows that the the occurrence of FSI leads to a redistribution of 
the strength dictated by the folding function $f_q(E)$, whose shape is 
strongly affected by nucleon-nucleon correlations \cite{benhar91}. In absence of FSI 
$f_q(E) \rightarrow \delta(E)$ and the IA cross section is recovered.

%%%%%%%%%%%%%%%%%%%%%%%%%%%%%%%%%%%%%%%%%%%%%%%%%%%%%%%%%%%%%%%%%%%%%%%%%%%%%%%%%%%%%%%%%%%%
\section{Results}
%%%%%%%%%%%%%%%%%%%%%%%%%%%%%%%%%%%%%%%%%%%%%%%%%%%%%%%%%%%%%%%%%%%%%%%%%%%%%%%%%%%%%%%%%%%%

We have computed the cross section of the quasielastic process
\beq
\nu_e + ^{16}O \rightarrow e^- + p + X\ ,
\label{process2}
\eeq
where $X$ denotes the recoiling spectator system. 
The spectral function apperaing in Eq.(\ref{tensor:IA}) has been obtained within 
the Local Density Approximation (LDA) \cite{benhar94}, while for the Dirac, Pauli and 
axial form factors entering the definition of the nucleon tensor $w^{\mu\nu}$ we have 
used the simple dipole parametrization.

The folding function has been evaluated using a parametrization of the measured 
nucleon-nucleon scattering amplitude \cite{oneill} and Monte Carlo configurations sampled 
from the probability distribution associated with the realistic oxygen wave function 
of ref. \cite{pieper92}.
%%%%%%%%%%%%%%%%%%%%%%%%%%%%%%%%%%%%%%%%%%%%%%%%%%%%%%%%%%%%%%%%%%%%%%%%%%%%
\begin{figure}[hbt]
\includegraphics[scale=.45]{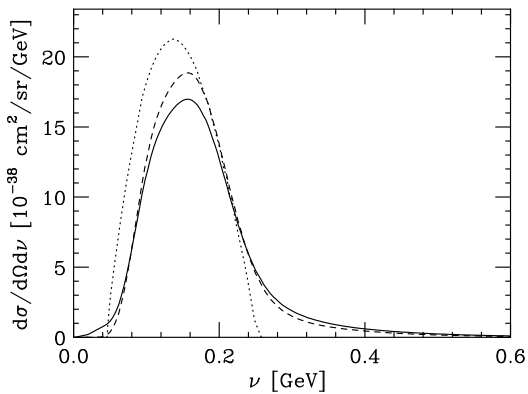}
\vspace*{-.3in}
\caption{\small
Differential cross section $d\sigma/d\Omega_e dE_e$
for neutrino energy $E = 1 $ GeV
and electron scattering angle $\theta_e = 30^\circ$.
The IA results are represented by the dashed line, while the solid line
correspond to the full calculation, including the effects of FSI.
The dotted line shows the prediction of the FG model
with Fermi momentum $k_F = 221$ MeV and average separation energy
$\epsilon = 35$ MeV.
}
\label{fig:1}
\end{figure}
%%%%%%%%%%%%%%%%%%%%%%%%%%%%%%%%%%%%%%%%%%%%%%%%%%%%%%%%%%%%%%%%%%%%%%%%%%%%%%

Fig. \ref{fig:1} shows the cross section corresponding to neutrino energy $E = 1 $ GeV, 
and electron scattering angle $\theta_e = 30^\circ$, plotted as a function of the 
energy transfer $E - E_e$. Comparison between the solid and dashed lines shows that 
the inclusion of FSI results in a sizable redistribution of the IA strength, 
leading to a quenching of the quasielastic peak and to the enhancement of the tails.
For reference, we also show the cross section predicted by the Fermi 
gas (FG) model with Fermi momentum $k_F = 221$ MeV and average separation energy
$\epsilon = 35$ MeV. It appears that nuclear dynamics, neglected in the oversimplified
picture in terms of noninteracting nucleons, plays a relevant role.

It has to be pointed out that the approach of ref. \cite{benhar91}, while including
dynamical correlations in the final state, does not take into account statistical
correlations, leading to Pauli blocking of the phase-space available to the 
outgoing nucleon. 

A rather crude prescription to include the effect of Pauli blocking amounts to 
modifying the spectral function according to
\beq
P(p) \rightarrow P(p) \theta(|{\bf p} + {\bf q}| - k_F)
\label{pauli}
\eeq
where $k_F$ is the average nuclear Fermi momentum, defined as
\label{local:kF}
\beq
k_F = \int  d^3r\ \rho_A({\bf r}) k_F({\bf r}), 
\eeq
with $k_F({\bf r})=(3 \pi^2 \rho_A({\bf r})/2 )^{1/3}$, $\rho_A({\bf r})$ being the 
measured nuclear density distribution. Eq.(\ref{local:kF}) yields $k_F = 209$ MeV. 
Note that, unlike the spectral function, the quantity defined in Eq.(\ref{pauli})
does not describe intrinsic properties of the target only, as it depends explicitely upon 
the momentum transfer.
%%%%%%%%%%%%%%%%%%%%%%%%%%%%%%%%%%%%%%%%%%%%%%%%%%%%%%%%%%%%%%%%%%%%%%%%%%%%
\begin{figure}[hbt]
\includegraphics[scale=.45]{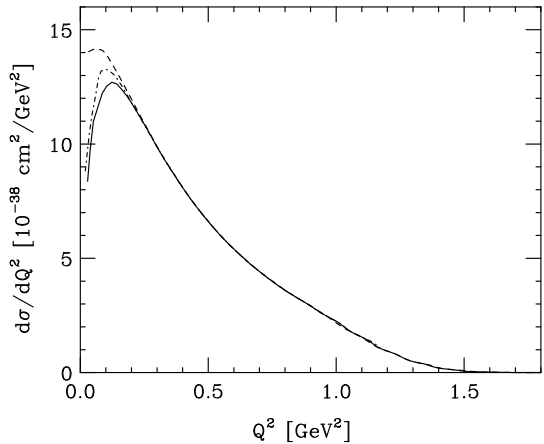}
\vspace*{-.3in}
\caption{\small
Differential cross section $d\sigma/dQ^2$
for neutrino energy $E= 1$ GeV. The dot-dash line shows the IA results, 
while the solid and dashed lines have been obtained using the modified 
spectral function ef Eq.(\protect\ref{pauli}), with and without inclusion 
of FSI, respectively.
}
\label{fig:2}
\end{figure}
%%%%%%%%%%%%%%%%%%%%%%%%%%%%%%%%%%%%%%%%%%%%%%%%%%%%%%%%%%%%%%%%%%%%%%%%%%%%%%

The effect of Pauli blocking is hardly visible in the differential cross sections
shown in Fig. \ref{fig:1}, as the kinematical setup corresponds to
$Q^2 \sim 2.4$ GeV$^2$ at the quasielastic peak. On the other hand, it becomes very
large at lower $Q^2$.

Fig. \ref{fig:2} shows the calculated differential cross section $d\sigma/dQ^2$
for neutrino energy $E= 1$ GeV. The dot-dash and dashed lines correspond to the
IA results with and without inclusion of Pauli blocking, respectively. It 
clearly appears that the effect of Fermi statistic in suppressing scattering 
at $Q^2 < .2$ GeV$^2$ is very large and must be taken into account. The results of
the full calculation, in which FSI are also included, are displayed as a full line. 
The results of Fig. \ref{fig:2} suggest that Pauli blocking and FSI may explain 
the deficit of the measured cross section at low $Q^2$ with respect to the 
predictions of Monte Carlo simulations \cite{Q2}.

\begin{figure}[hbt]
\includegraphics[scale=.45]{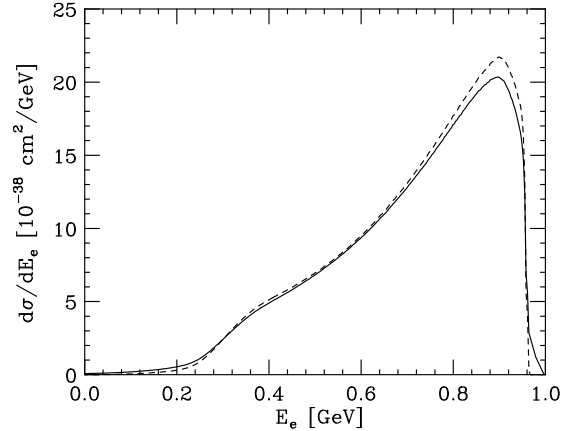}
\vspace*{-.3in}
\caption{\small
Energy spectrum of the produced electrons in charged current interactions at 
neutrino energy $E = 1$ GeV. The solid and dashed lines have been obtained including 
the effect of Pauli blocking with and without FSI, respectively.
}
\label{fig:3}
\end{figure}

The energy spectrum of the produced electrons obtained using the modified 
spectral function of Eq.(\ref{pauli}) is displayed in Fig. \ref{fig:3}. 
The solid and dashed lines corresponds to the results of calculations 
carried out with and without FSI.
As in Fig. \ref{fig:1}, it appears that inclusion of FSI results in a 
redistribution of the strength, leading to a quenching 
in the region of the peak and to the appearance of a tail at large energy, 
extending all the way to the boundary of the kinematically allowed region.
%%%%%%%%%%%%%%%%%%%%%%%%%%%%%%%%%%%%%%%%%%%%%%%%%%%%%%%%%%%%%%%%%%%%%%%%%%%%
%%%%%%%%%%%%%%%%%%%%%%%%%%%%%%%%%%%%%%%%%%%%%%%%%%%%%%%%%%%%%%%%%%%%%%%%%%%%%%
\section{Conclusions}
%%%%%%%%%%%%%%%%%%%%%%%%%%%%%%%%%%%%%%%%%%%%%%%%%%%%%%%%%%%%%%%%%%%%%%%%%%%%%%%%%%%%%%%%%%%%

NMBT provides a parameter free and computationally viable framework to study the
nuclear response to weakly interacting probes, whose quantitative understanding will be 
critical to the analysis of the next genaration of high precision neutrino oscillation 
experiments.

In the IA scheme, which is likely to be applicable to describe scattering of few GeV
neutrinos, NMBT allows for a consistent treatment of dynamical nucleon-nucleon 
correlations, that largely affect both the nucleon spectral function and FSI. 
Electron scattering experiments have provided overwhelming evidence that the independent 
particle model fails to explain the data and inclusion of correlation effects,  
leading to a sizable redistribution of the strength, is indeed required.

As a final remark, it has to be emphasized that
the extension of the approach described in this paper to include exclusive as well
as inelastic channels, along the lines of the work done for electron-nucleus 
scattering, is straightforward.

%%%%%%%%%%%%%%%%%%%%%%%%%%%%%%%%%%%%%%%%%%%%%%%%%%%%%%%%%%%%%%%%%%%%%%%%%%%%%%%%%%%%%%%%%%%

\end{document}